**How to design a pre-specified statistical analysis approach to limit p-hacking in clinical trials: the Pre-SPEC framework**

Brennan C Kahan[1]*, Gordon Forbes[2], Suzie Cro[3]

[1] MRC Clinical Trials Unit at UCL, 90 High Holborn, London, WC1V 6LJ

[2] Department of Biostatistics and health informatics, Institute of Psychiatry, Psychology & Neuroscience, Kings College London, 16 De Crespigny Park, London, SE5 8AF

[3] Imperial Clinical Trials Unit, Imperial College London, Stadium House, 68 Wood Lane, London, W12 7RH

*Brennan Kahan: b.kahan@ucl.ac.uk

**Abstract**

Results from clinical trials can be susceptible to bias if investigators choose their analysis approach after seeing trial data, as this can allow them to perform multiple analyses and then choose the method that provides the most favourable result (commonly referred to as 'p-hacking'). Pre-specification of the planned analysis approach is essential to help reduce such bias, as it ensures analytical methods are chosen in advance of seeing the trial data. For this reason, guidelines such as SPIRIT (Standard Protocol Items: Recommendations for Interventional Trials) and ICH-E9 (International Conference for Harmonisation of Technical Requirements for Pharmaceuticals for Human Use) require the statistical methods for a trial's primary outcome be pre-specified in the trial protocol. However, pre-specification is only effective if done in a way that does not allow p-hacking. For example, investigators may pre-specify a certain statistical method such as multiple imputation, but give little detail on how it will be implemented. Because there are many different ways to perform multiple imputation, this approach to pre-specification is ineffective, as it still allows investigators to analyse the data in different ways before deciding on a final approach. In this article we describe a five-point framework (the Pre-SPEC framework) for designing a pre-specified analysis approach that does not allow p-hacking. This framework was designed based on the principles in the SPIRIT and ICH-E9 guidelines, and is intended to be used in conjunction with these guidelines to help investigators design the statistical analysis strategy for the trial's primary outcome in the trial protocol.

**Introduction**

Results from clinical trials depend upon the statistical methods used for analysis (1-5). Different methods of analysis applied to the same trial can lead to different conclusions around effectiveness and safety (1-14). Therefore, results from clinical trials can be susceptible to bias if investigators choose their analysis approach after seeing trial data, as this can allow them to perform multiple analyses and then choose the approach that provides the most favourable result. This is commonly referred to as 'p-hacking', and can lead to bias in treatment effect estimates, confidence intervals, and p-values (1-5, 7-10, 12, 15). Pre-specification of the planned analysis approach is therefore essential to help reduce such bias, as it ensures that analytical methods are chosen in advance of seeing the trial data (1-5, 7, 9, 10, 12). The SPIRIT (*Standard Protocol Items: Recommendations for Interventional Trials*) and ICH-E9 (*International Conference for Harmonisation of Technical Requirements for Pharmaceuticals for Human Use*) guidelines require that the method of analysis for the trial's primary outcome be pre-specified in the trial protocol (1, 3, 4).

However, pre-specification is only effective if done in a way that does not allow p-hacking. For example, investigators may pre-specify a certain statistical method, such as multiple imputation to handle missing data, but give little detail on how it will be implemented. However, there are many different ways to implement multiple imputation, such as including different variables in the imputation model, imputing under different statistical models, etc. Therefore, this approach to pre-specification is ineffective, as it still allows investigators to analyse the data in many different ways before deciding on a final approach. This issue of 'incomplete' pre-specification, where methods are pre-specified to some extent but the specification still allows for some degree of p-hacking, is common in clinical trials (table 1) (2-5). For example, two reviews which examined trial protocols found that 11-20% of protocols did not specify the analysis model that would be used for the primary outcome, 42% did specify the model but omitted

4essential detail on how the model would be implemented, and 19% specified an approach that would allow the investigators to subjectively choose the final analysis model after seeing the trial data (2, 5).

The SPIRIT and ICH-E9 documents contain guidance on what statistical content should be included in the trial protocol (1, 4), and there are also guidelines for the content of Statistical Analysis Plans (9). These guidance documents contain some statistical principles which help to limit p-hacking (e.g. requiring that when multiple analysis strategies are planned, one of them is identified as the primary analysis), however the primary aim of these guidelines is to describe *what* information should be included in the protocol or Statistical Analysis Plan, rather than describe exactly *how* the analysis should be designed. As such, these guidelines do not offer a prescriptive approach for how analysis strategies should be designed in order to limit p-hacking. In this article, we describe a framework for how a statistical analysis strategy could be designed to ensure it does not allow p-hacking (i.e. so that no part of the statistical methods can be chosen after seeing the trial data in order to 'improve' results) (2-4). This framework was developed to be consistent with the statistical principles outlined in the SPIRIT and ICH-E9 guidelines (a comparison is shown in Additional file 1: Table S1), and is intended to be used in conjunction with these guidelines (1, 3, 4) to help investigators design the statistical analysis strategy for the trial's primary outcome in the trial protocol.

**The Pre-SPEC framework**

We now outline the Pre-SPEC framework (box 1). The five points are: (1) Pre-specify before recruitment to the trial begins; (2) Specify a single primary analysis strategy; (3) Plan each aspect of the analysis; (4) Enough detail should be provided so that a third party could independently perform the analysis; and (5) Adaptive analysis strategies should use deterministic decision rules. We expand on each of these points below.

***Pre-specify the analysis strategy before recruitment to the trial begins***

Pre-specifying the analysis strategy before the trial begins ensures the choice of methods is not influenced by any trial data. This can give readers confidence that trial results are not due to p-hacking (1, 3, 4), as they will generally have no way to verify that analyses specified after the trial began were not based on trial data.

Pre-specifying the analysis approach for the trial's primary outcome in the protocol before the trial begins is a requirement of both the SPIRIT and ICH-E9 guidelines (see Additional file 1: Table S1). For instance, ICH-E9 states that *"… the principal features of its proposed statistical analysis should be clearly specified in a protocol written before the trial begins"*, and *"… the principal features of the eventual statistical analysis of the data should be described in the statistical section of the protocol. This section should include all the principal features of the proposed confirmatory analysis of the primary variable(s) and the way in which anticipated analysis problems will be handled"* (1), while SPIRIT states *"The planned methods of statistical analysis should be fully described in the protocol"* and *"The protocol should indicate explicitly each intended analysis comparing study groups. An unambiguous, complete, and transparent description of statistical methods facilitates execution, replication, critical appraisal, and the ability to track any changes from the original pre-specified methods"* (4).

***Specify a single primary analysis strategy***

Specifying a single primary analysis strategy ensures investigators cannot perform multiple analyses and then selectively report the most favourable as their main approach. There are often valid reasons to specify additional methods of analysis, for instance to answer different questions about the intervention (e.g. the effect of a treatment policy vs the effect if everyone adheres (16)), or to assess the robustness of the main results to different assumptions about the data (e.g. sensitivity analyses for missing data (17)). In these instances, a single approach should be clearly labelled as the primary analysis strategy, with other approaches identified as sensitivity or supplementary analyses as appropriate (1, 3, 4).

*Plan each aspect of the statistical analysis*

Omission of a particular aspect from the analysis strategy could allow investigators to run multiple analyses for that aspect, and selectively report the most favourable. For example, if the analysis population is not specified, investigators could run both an intention-to-treat and per-protocol analysis, and present whichever is most favourable.

The minimum set of essential aspects to cover are:
- Analysis population
- Statistical model
- The use of covariates
- Handling of missing data

However, for many trials there will be additional aspects to cover; for instance, a non-inferiority trial would need to specify the non-inferiority margin.

It is also useful to specify the target estimand (16) and what information will be presented from the analysis, such as the level of the confidence interval and the threshold for statistical significance if applicable.

***Enough detail should be provided so that a third party could independently perform the analysis***

There is often a substantial amount of detail required to implement an analysis. For example, using multiple imputation for missing data requires specification of the method of imputing data; this includes specifying which variables are included in the imputation model (and how they are included), whether multivariate normal, chained equations or some other imputation approach is used, the number of imputed datasets to be used, and how imputed datasets will be combined. Simply stating that multiple imputation will be used is not sufficient, as this allows the investigator to carry out multiple analyses based on different imputation approaches, each of which could give a different result.

Fully pre-specifying these details to such a degree that a third party could independently perform the analysis helps to ensure investigators cannot perform multiple analyses. A good test of whether there is sufficient detail is to write out the statistical code that would be used to implement the analysis in a statistical software program; if investigators are unable to write out their planned code, this likely means the analysis strategy is not sufficiently well specified. This code could be tested on a simulated (fake) dataset to ensure if performs as intended.

An additional benefit to providing this code as a supplement to the description of the planned analysis in the protocol is that it leaves no room for ambiguity, and ensures all necessary detail is provided (18)**.**

***Adaptive analysis strategies should use deterministic decision rules***

Sometimes investigators use adaptive analysis strategies, where some aspect of the final analysis is chosen based on the trial data. For instance, they may specify that either multiple imputation or a complete case analysis will be used depending on the level of missing data. Many clinical trials will not require such decision rules, as there will often be an available analysis approach which can provide valid results under minimal assumptions about the data. However, investigators may find these rules useful in certain settings where their preferred approach will depend on some features of the data, which are not known in advance.

Adaptive analysis strategies can be problematic if the decision rules are subjective, as this allows investigators to perform each potential analysis and selectively report the most favourable. For example, without a clear rule about when to use multiple imputation vs. complete cases, investigators could perform both and then select whichever gives a 'better' result.

In order to prevent decisions from being driven by results, adaptive analysis strategies should use deterministic decision rules for selection of the final analysis approach. A decision rule is deterministic if two different people are guaranteed to get the exact same result by following the rule. This removes the investigators ability to influence decisions, and will therefore ensure results cannot be p-hacked. In the example above, investigators could specify that multiple imputation will be used if the level of missing outcome data is >5%, and a complete case analysis will be used otherwise.

We note that in many instances adaptive analysis strategies can lead to biased estimates or incorrect standard errors even when decision rules are fully deterministic. For example, this occurs when using stepwise selection to choose which covariates to adjust for; when using a test for carryover to determine the final analysis model in a crossover design; or when using a test for interaction to determine the final analysis model for a factorial trial (19-21). Therefore, caution should be applied when considering adaptive strategies, even if deterministic decision rules are planned.

**Example**

We now illustrate our framework in an example. Consider the following analysis section from a trial protocol for a continuous primary outcome measured at multiple follow-up time-points:

*"Primary analyses will be undertaken on an intention-to-treat basis, including all participants as randomised, regardless of treatment actually received. The intervention group will be compared with the control group using a planned contrast of change from baseline to the week 12 endpoint using a mixed-model repeated measures analysis. Stratification variables will be evaluated and retained in analyses where they are measured as significant or quasi-significant. Transformation of outcomes, including categorisation, may be undertaken to meet distributional assumptions and accommodate outliers."*

*Evaluating whether the analysis approach is designed to prevent p-hacking*

This analysis approach meets our first two points; it was described in the trial protocol before recruitment began, and consists of a single overall analysis strategy.

For our third point, the analysis approach covers three analysis aspects (population, analysis model, covariates), however it does not specify how missing data will be handled. We can guess that participants with missing outcome data at all follow-up time-points will be excluded from the analysis, however this is not entirely clear.

For our fourth point, there is insufficient detail for a third party to independently replicate the analysis model; there are numerous ways to implement a mixed-model repeated measures analysis (for instance, different approaches to specifying random-effects, or different correlation structures to model the correlation between outcomes from the same participant at different time points), and it is not clear which approach the authors intend to use.

For our fifth point, the authors plan to use an adaptive analysis strategy for two components; which stratification variables to include in the analysis, and whether to transform the outcome (and if so, which transformation to use). In both instances, they do not include deterministic decision rules on how the final analysis approach should be decided (e.g. for stratification variables, there is no definition of what quasi-significant means). Therefore, this strategy would allow investigators to perform multiple analyses on the final trial data before choosing their preferred approach.

Overall, the specified analysis approach could allow investigators to implement a number of different analysis strategies (relating to handling of missing data, the analysis model, covariates, and transformation of the outcome) and present the most favourable result. As such, although this approach has been pre-specified, it still allows p-hacking.

***Modifying the analysis approach so it is designed to prevent p-hacking using the Pre-SPEC framework***

We can modify the approach described in the previous section so that it does not allow p-hacking by resolving the issues relating to points 3-5 above. First, we could explicitly state that the analysis will use all available follow-up data; participants with an available outcome from at least one follow-up time point will be included in the analysis, and participants with missing outcome data at all follow-up time points will be excluded from the analysis.

Second, we could provide additional information on how the analysis model will be implemented; for instance, we could specify a linear mixed-effects model with an unstructured correlation matrix for observations at different time-points, estimated using restricted maximum likelihood. We could supplement this description by including the planned statistical code to remove any ambiguity from our description (see below for example code for the statistical package *Stata*).

Finally, we need to resolve the issues around the adaptive analysis strategies related to the stratification variables and the transformation of the outcome. In this scenario, it is unlikely that the adaptive strategies are necessary, or even beneficial. All stratification variables should be included in the model regardless of statistical significance, as failure to do so can lead to incorrect confidence intervals and p-values (22, 23). Furthermore, linear regression models are usually very robust to violations of distributional assumptions (24), and transformation can lead to issues of interpretability (in particular, categorisation could lead to a substantial reduction in power (25)). Therefore, the simplest way to resolve this issue is to remove the adaptive part, and use a strategy which includes all stratification variables in the model and does not consider transformations of the outcome. This approach would guarantee valid results under minimal assumptions about the data, which are easily interpretable. If an adaptive strategy was deemed necessary, then a deterministic decision rule would need to be specified, for example by giving the exact p-value threshold for retaining stratification variables in the model (though we note this approach can be problematic even if fully pre-specified (21)).

Incorporating these changes, we could re-write the planned analysis strategy as follows:

*Primary analyses will be undertaken on an intention-to-treat basis, including all participants as randomised, regardless of treatment actually received. The analysis will use all available outcome data; participants with an available outcome from at least one follow-up time point will be included in the analysis, and participants with no recorded outcomes will be excluded from the analysis. The intervention*

*group will be compared with the control group using a planned contrast of change from baseline to the week 12 endpoint and will be fit using a linear mixed-model which includes outcomes at all time-points in the model. The model will use an unstructured correlation matrix for observations at different time points, and will be fit using restricted maximum likelihood. The model will include treatment group, time point, a treatment-by-time interaction, and the stratification variables as fixed factors. This analysis will be implemented using the following Stata code:*

```
mixed outcome treat_group i.time_point treat_group#i.time_point strat1
strat2 || patient_id:, res(unstructured, t(time_point)) noconstant
reml
```

*lincom treat_group+treat_group#12.time_point*

*Where 'outcome' refers to the primary outcome (change from baseline), 'treat_group' to the treatment group, 'time_point' refers to the follow-up time-point, 'treat_group#i.time_point' refers to the treatment group by follow-up time-point interaction, 'strat1' and 'strat2' refer to the stratification variables and 'participant_id' is a unique ID for participant. The treatment effect at week 12 (primary outcome) is estimated using the Stata code: lincom treat_group+treat_group#12.time_point*

We note that Stata automatically excludes participants with no recorded outcomes from the analysis, and so does not require additional code to perform this step. Further, we note that the above strategy is not necessarily the optimal statistical approach, but is used simply to illustrate how the original approach could be fully pre-specified.

**Discussion**

Pre-specification of the planned statistical analysis approach can help to help reduce bias from p-hacking in clinical trials, as it ensures analytical methods are chosen in advance of seeing the trial data. However, 'incomplete' pre-specification, which still allows some degree of p-hacking, is common in clinical trials (2, 5). Pre-SPEC is a framework that describes how a statistical analysis strategy could be designed to ensure it does not allow p-hacking.

This framework was designed to be consistent with the SPIRIT and ICH-E9 guidelines (1, 4), and is intended to be used in conjunction with these and other guidelines (9). The SPIRIT and ICH-E9 guidelines require the analysis strategy for a trial's primary outcome be documented in the trial protocol, and as such, the Pre-SPEC framework is intended to help investigators design the analysis strategy for the trial's primary outcome in the trial protocol. Our intention is not for the use of this framework be mandated, but rather for it to provide guidance for those who wish to design a statistical analysis approach which both (i) does not allow p-hacking; and (ii) can be seen by others to not allow p-hacking.

The statistical analysis approach for the trial's primary outcome is usually specified well in advance of the trial start date, as it is often required for grant application or the sample size calculation. Therefore, this information will usually be available to include in the trial protocol. However, for trials for which this information is not known at the protocol stage, and where investigators feel that specifying this information would pose an insurmountable barrier to the timely start of the trial, then investigators should specify the planned analysis approach for the primary outcome as soon after the trial has begun as possible. For these trials, it may be difficult for readers to determine whether the planned analysis approach was specified before investigators had access to unblinded trial data, and so accurate reporting around when trial investigators and statisticians received data, and whether they were blinded to treatment allocation codes within the dataset, is essential to allow transparent evaluation of results (26, 27).

Although this framework was developed with a trial's primary outcome in mind, it could also be used for secondary outcomes. As above, where investigators feel that specifying this information would pose an insurmountable barrier to the timely start of the trial, then investigators should simply specify the planned analysis approach as soon after the trial has begun as possible. Importantly, we note that our framework does not require that a detailed Statistical Analysis Plan be written before the trial begins.

We note that the Pre-SPEC framework is not intended to preclude changes, or force investigators to stick with an analysis strategy they feel is no longer appropriate. There are sometimes good reasons for investigators to change their statistical methods during the course of the trial, for instance because of an advance in statistical methodology or the implementation of new methods in statistical software packages. Instead, if it is anticipated beforehand that the preferred method of analysis may depend on some aspect of the trial data (for instance, the distribution of outcome data), then the manner in which this decision will be made should be pre-specified; and, if the analysis strategy needs to change due to an unanticipated issue (for instance, the occurrence of unanticipated intercurrent events (28), or new methodology becoming available in statistical software packages), then these changes should be documented and explained (26). Instead of preventing useful or necessary changes, Pre-SPEC simply increases transparency around the process; as stated in the SPIRIT guidelines, "*An unambiguous, complete, and transparent description of statistical methods facilitates execution, replication, critical appraisal, and the ability to track any changes from the original pre-specified methods.*" (4)

15We note that transparency around the statistical methods used in clinical trials is increasing, and there are initiatives in place to further increase transparency (for example, those conducted by the UKCRC CTU network, https://www.ukcrc-ctu.org.uk/). However, there is still a long way to go; evidence shows that the statistical methods for the trial's primary outcome are often poorly specified in both trial protocols (5, 26, 27) and Statistical Analysis Plans (26); that protocols and Statistical Analysis Plans are often not made publicly available, or are only done so after they may have already been modified during the course of the trial (26, 27, 29); undisclosed changes to the planned analysis approach are frequent (2, 26, 27); and reporting around data access and blinding status of statisticians is often poor (26, 27), hampering the ability of readers to evaluate whether changes have been made based on unblinded trial data. Pre-SPEC can play a part, alongside other initiatives, to help increase transparency in clinical trials, and resolving some of the issues outlined above.

**Conclusion**

Use of the Pre-SPEC framework can help ensure that statistical analyses are designed so they do not allow p-hacking.

**Table 1: Common issues in pre-specifying statistical analysis approaches in clinical trial protocols**

| Issue | Problems associated with issue | Estimated prevalence | |
|---|---|---|---|
| | | Aspect | Prevalence[a] |
| Omitting an aspect of the analysis approach | | Analysis population:<br>Analysis model:<br>Covariates:<br>Missing data: | 27-47%<br>11-20%<br>27%<br>66-77% |
| Insufficient detail around an aspect of the analysis approach | Investigators could run multiple analyses, and selectively report the most favourable | Analysis population:<br>Analysis model:<br>Covariates:<br>Missing data: | 64%<br>42%<br>23%<br>17%[b] |
| Analysis approach allows some aspects of the final analysis to be subjectively chosen based on trial data | | Analysis model:<br>Covariates: | 19%<br>8% |
| Multiple analysis approaches specified, without one being identified as the primary | Investigators could selectively report the most favourable result, or to elevate its importance compared to less favourable results. | Analysis population:<br>Analysis model:<br>Covariates:<br>Missing data: | 11%<br>11%<br>9%<br>2% |

[a] Based on references (5) and (2); one study evaluated protocols and published results for 70 randomised trials approved by the ethics committees for Copenhagen and Frederiksberg, Denmark in 1994-5; the other study evaluated 100 protocols of randomised trials indexed in PubMed November 2016.

[b] 15/99 protocols gave insufficient detail around how they planned to implement multiple imputation, 2/99 protocols but gave insufficient detail around their planned inverse probability weighting procedure

**Box 1 – Framework for pre-specifying a statistical analysis strategy (Pre-SPEC)**

| | |
|---|---|
| **Pre**-specify before recruitment | Pre-specify the analysis strategy before recruitment to the trial begins. |
| **S**ingle analysis strategy | Specify a single primary analysis strategy. |
| **P**lan each aspect | Each aspect of the planned analysis should be covered, including analysis population, statistical model, covariates, and handling of missing data. |
| **E**nough detail | Provide sufficient detail to allow a third party to independently perform the analysis (ideally through statistical code). |
| **C**hoices made deterministically | For adaptive analysis strategies which use the trial data to inform some aspect of the analysis, use deterministic decision-rules that prevent analysis choices being driven by results. |

## Declarations

**Ethics approval and consent to participate**

Not applicable

**Consent for publication**

Not applicable

**Availability of data and materials**

Not applicable

**Competing interests**

The authors declare that they have no competing interests.

**Funding**

None.

**Authors' contributions**

BK conceived the idea for this article and wrote the first draft. GF and SC contributed to the manuscript and helped refine the Pre-SPEC framework. All authors read and approved the final manuscript.

**Acknowledgements**

We would like to thank Thomas Bandholm, Victoria Cornelius, Rachel Phillips, Francesca Fiorentino, Nicholas Johnson, Consuelo Nohpal de la Rosa and Jinky Lozano-Kuehne for helpful comments on a draft of the manuscript.

**Additional file 1: Table S1**

**Table S1 – Comparison of the Pre-SPEC framework with the SPIRIT and ICH-E9 guidelines**

| Pre-SPEC framework | SPIRIT | ICH-E9 | Comment |
|---|---|---|---|
| Pre-specify before recruitment to the trial begins | • "The planned methods of statistical analysis should be fully described in the protocol"<br>• "The protocol should indicate explicitly each intended analysis comparing study groups. An unambiguous, complete, and transparent description of statistical methods facilitates execution, replication, critical appraisal, and the ability to track any changes from the original pre-specified methods." | • "For each clinical trial contributing to a marketing application, all important details of its design and conduct and the principal features of its proposed statistical analysis should be clearly specified in a protocol written before the trial begins." (p5)<br>• "When designing a clinical trial the principal features of the eventual statistical analysis of the data should be described in the statistical section of the protocol. This section should include all the principal features of the proposed confirmatory analysis of the primary variable(s) and the way in which anticipated analysis problems will be handled." p23-24 | Both SPIRIT and ICH-E9 state the planned statistical analysis approach should be pre-specified in the protocol. ICH-E9 explicitly states this should be done before the trial begins; SPIRIT does not state this explicitly, but it is implied given that the first version of the protocol must be completed before the trial begins. |
| Specify a single primary analysis strategy. | • "Results for the primary outcome can be substantially affected by the choice of analysis methods. When investigators apply more than one analysis strategy for a specified primary outcome, there is potential for inappropriate selective reporting of the most | • "The primary analysis of the primary variable should be clearly distinguished from supporting analyses of the primary or secondary variables." p28 | Both SPIRIT and ICH-E9 state explicitly that a single main analysis strategy should be identified. |
| | interesting result. The protocol should prespecify the main ("primary") analysis of the primary outcome…"<br>• "When both unadjusted and adjusted analyses are intended, the main analysis should be identified (Item 20a)." | | |
| Plan all aspects of the analysis (including analysis population, statistical model, covariates, and handling of missing data) | • "The protocol should prespecify the main ("primary") analysis of the primary outcome (Item 12), including the analysis methods to be used for statistical comparisons (Items 20a and 20b); precisely which trial participants will be included (Item 20c); and how missing data will be handled (Item 20c)."<br>• "It is important that trial investigators indicate in the protocol if there is an intention to perform or consider adjusted analyses, explicitly specifying any variables for adjustment and how continuous variables will be handled."<br>• "Protocols should explicitly describe which participants will be included in the main analyses (eg, all randomised | • "The set of subjects whose data are to be included in the main analyses should be defined in the statistical section of the protocol." p24<br>• "The decision to transform key variables prior to analysis is best made during the design of the trial on the basis of similar data from earlier clinical trials. Transformations (e.g. square root, logarithm) should be specified in the protocol and a rationale provided, especially for the primary variable(s)." p27<br>• "The statistical section of the protocol should specify the hypotheses that are to be tested and/or the treatment effects which are to be estimated in order to satisfy the primary objectives of the trial. The statistical methods to be used to accomplish these tasks should be described for the primary (and preferably the secondary) variables, and the underlying | SPIRIT explicitly states that the analysis population, analysis model, covariates, handling of missing data, and any other relevant aspects should be specified. ICH-E9 explicitly states that the analysis population, statistical model, covariates, and use of transformations for key variables should be specified. |
| | participants, regardless of protocol adherence) and define the study group in which they will be analysed (eg, as randomised)."<br>• "The protocol should also state how missing data will be handled in the analysis and detail any planned methods to impute (estimate) missing outcome data, including which variables will be used in the imputation process (if applicable)."<br>• "Finally, different trial designs dictate the most appropriate analysis plan and any additional relevant information that should be included in the protocol. For example, cluster, factorial, crossover, and within-person randomised trials require specified statistical considerations, such as how clustering will be handled in a cluster randomised trial." | statistical model should be made clear. Estimates of treatment effects should be accompanied by confidence intervals, whenever possible, and the way in which these will be calculated should be identified. A description should be given of any intentions to use baseline data to improve precision or to adjust estimates for potential baseline differences, for example by means of analysis of covariance." p27<br>• "All effects to be fitted in the analysis (for example in analysis of variance models) should be fully specified… . The same considerations apply to the set of covariates fitted in an analysis of covariance." p28<br>• "The primary variable(s) is often systematically related to other influences apart from treatment. For example, there may be relationships to covariates such as age and sex, or there may be differences between specific subgroups of subjects such as those treated at the different centres of a multicentre trial. In some instances an adjustment for the influence of covariates or for subgroup effects is an integral part of the planned analysis and hence should be set | |
| | | out in the protocol. Pre-trial deliberations should identify those covariates and factors expected to have an important influence on the primary variable(s), and should consider how to account for these in the analysis in order to improve precision and to compensate for any lack of balance between treatment groups." p28 | |
| Enough detail should be provided so that a third party could independently perform the analysis | • "It is important that trial investigators indicate in the protocol if there is an intention to perform or consider adjusted analyses, explicitly specifying any variables for adjustment and how continuous variables will be handled."<br>• "Protocols should explicitly describe which participants will be included in the main analyses (eg, all randomised participants, regardless of protocol adherence) and define the study group in which they will be analysed (eg, as randomised)."<br>• "The ambiguous use of labels such as "intention to treat" or "per protocol" should be avoided unless they are fully defined in the protocol. … | • "The decision to transform key variables prior to analysis is best made during the design of the trial on the basis of similar data from earlier clinical trials. Transformations (e.g. square root, logarithm) should be specified in the protocol and a rationale provided, especially for the primary variable(s)." p27<br>• "The statistical section of the protocol should specify the hypotheses that are to be tested and/or the treatment effects which are to be estimated in order to satisfy the primary objectives of the trial. The statistical methods to be used to accomplish these tasks should be described for the primary (and preferably the secondary) variables, and the underlying statistical model should be made clear. Estimates of treatment effects should be accompanied by | Both SPIRIT and ICH-E9 state that certain aspects of the analysis should be explicitly or fully described (e.g. analysis population, covariates, handling of missing data, etc). |

| | | |
|---|---|---|
| | Other ambiguous labels such as "modified intention to treat" are also variably defined from one trial to another."<br>• "The protocol should also state how missing data will be handled in the analysis and detail any planned methods to impute (estimate) missing outcome data, including which variables will be used in the imputation process (if applicable)." | confidence intervals, whenever possible, and the way in which these will be calculated should be identified. A description should be given of any intentions to use baseline data to improve precision or to adjust estimates for potential baseline differences, for example by means of analysis of covariance." p27<br>• "All effects to be fitted in the analysis (for example in analysis of variance models) should be fully specified… . The same considerations apply to the set of covariates fitted in an analysis of covariance." p28<br>• "The primary variable(s) is often systematically related to other influences apart from treatment. For example, there may be relationships to covariates such as age and sex, or there may be differences between specific subgroups of subjects such as those treated at the different centres of a multicentre trial. In some instances an adjustment for the influence of covariates or for subgroup effects is an integral part of the planned analysis and hence should be set out in the protocol. Pre-trial deliberations should identify those covariates and factors expected to | |
| | | have an important influence on the primary variable(s), and should consider how to account for these in the analysis in order to improve precision and to compensate for any lack of balance between treatment groups." p28 | |
| Adaptive analysis strategies should use deterministic decision rules | • "It is important that trial investigators indicate in the protocol if there is an intention to perform or consider adjusted analyses… . It may not always be clear, in advance, which variables will be important for adjustment. In such situations, the objective criteria to be used to select variables should be prespecified." | • "The particular statistical model chosen should reflect the current state of medical and statistical knowledge about the variables to be analysed as well as the statistical design of the trial. All effects to be fitted in the analysis (for example in analysis of variance models) should be fully specified, and the manner, if any, in which this set of effects might be modified in response to preliminary results should be explained." p28 | SPIRIT advocates objective decision rules in a single specific instance (if covariates are to be chosen based on trial data). ICH-E9 states that the way the analysis might be modified in response to preliminary results should be specified. To the extent that adaptive analysis strategies are mentioned, both imply that pre-specified deterministic decision rules should be used. |